\documentclass[onecolumn,letterpaper, 12 pt,journal]{ieeeconf}

\IEEEoverridecommandlockouts
\usepackage{cite}
\usepackage{amsmath,amssymb,amsfonts}
\usepackage{algorithmic}
\usepackage{graphicx}
\usepackage{textcomp}
\usepackage{xcolor}
\def\BibTeX{{\rm B\kern-.05em{\sc i\kern-.025em b}\kern-.08em
    T\kern-.1667em\lower.7ex\hbox{E}\kern-.125emX}}
\title{\LARGE \bf
	Advanced control based on Recurrent Neural Networks learned using Virtual Reference Feedback Tuning and application to an Electronic Throttle Body (with supplementary material)
}
\author{William D'Amico, Marcello Farina, Giulio Panzani%
	\thanks{The authors are with Dipartimento di Elettronica, Informazione e Bioingegneria,
		Politecnico di Milano, Via Ponzio 34/5, 20133, Milano, Italy
		{\tt\small \{name.surname\}@polimi.it}}%
}
\begin{document}
%
%

\maketitle
\thispagestyle{empty}
\pagestyle{empty}

\begin{abstract}
In this paper the application of Virtual Reference Feedback Tuning (VRFT) for control of nonlinear systems with regulators defined by Echo State Networks (ESN) and Long Short Term Memory (LSTM) networks is investigated. The capability of this class of regulators of constraining the control variable is pointed out and an advanced control scheme that allows to achieve zero steady-state error is presented. The developed algorithms are validated on a benchmark example that consists of an electronic throttle body (ETB).
\end{abstract}
%

\section{Introduction}
Nowadays, data science is increasingly spreading due to the recent introduction of innovative tools and algorithms for extracting information from data \cite{b1}.

In the recent decades, data-based identification and control techniques \cite{b2} have been subject of increasing attention in research. In automation and control, several approaches are devoted to \textit{indirect} data-based methods for control design that aim at first identifying a model of the plant, based on which the controller is then designed \cite{b3}. The plant model is identified by means of linear structures, e.g. ARX (AutoRegressive models with eXogenous variables), or nonlinear ones, e.g. Hammerstein-Wiener models or Recurrent Neural Networks (RNN) \cite{b4}.

On the other hand, \textit{direct} data-based methods for control design provide a valid alternative to indirect methods. In this case, the controller is directly identified through optimization from a controller class previously selected without preliminarily using the data to identify a model of the plant. They are either based on a reference model, e.g. Virtual
Reference Feedback Tuning (VRFT) \cite{b5}, Iterative Learning \cite{b6}, or they exploit more recent model-free techniques, e.g. based on Reinforcement Learning \cite{b7}. 

In particular, VRFT has been first introduced for linear controller design \cite{b8} and then has been extended to nonlinear controller classes \cite{b9}. The use of VRFT to tune controllers of the class of Neural Networks (NN) has been marginally investigated. In \cite{b10} and \cite{b11}, the filter implementation in the training algorithm of RNNs using dynamic backpropagation of the gradients is discussed and applied to a simulated crane
model. In \cite{b12}, VRFT is applied to Multi-Input–Multi-Output (MIMO) nonlinear systems using a three-layer neural network controller trained by least squares. Finally, in \cite{b13}, this method is used for tuning fractional order (FO) and state-feedback NN controllers.

Among NNs, it has been proved that RNNs have promising capabilities in modeling and control of nonlinear systems \cite{b14}. Among RNNs, Echo State Networks (ESN) \cite{b15} and Long Short Term Memory (LSTM) networks \cite{b16} are particularly advantageous since they do not suffer from the vanishing gradient problem \cite{b17}, an issue occurring during the training of recurrent networks in case the gradients of the network output with respect to the parameters in the early layers become very small. Moreover, as discussed in \cite{b18} and in \cite{b19} (dealing with ESNs and LSTMs, respectively), some notable properties of these classes of RNNs (i.e., $\delta$ISS \cite{b19b}) can be easily studied and even enforced during the training phase.

In this paper, we focus on the application of VRFT for controller design, with attention to regulators with ESN and LSTM structures. 
The use of these networks has a twofold advantage: first, we can safely enforce stability-related properties to the controller; secondly, thanks to the nonlinear structures of these RNNs, we can explicitly enforce constraints on the control variables by constraining the networks parameters in the learning phase. The latter is a remarkable advantage over other commonly-used controller structures, since in most applications the control variables are bounded in defined ranges. Among direct methods, only in \cite{b20} an alternative is provided by means of a hierarchical control architecture, where an inner controller is first designed to match an a-priori closed-loop model and an outer model predictive controller is synthesized to consider input/output constraints and to enhance the performance of the inner loop.\\
In the advanced control scheme proposed in this paper, steady-state performance is guaranteed by suitable explicit integral actions. In this work, for notational simplicity (and in view of the fact that the considered case study has one input and one output), we consider single-input and single-output (SISO) networks, although a generalization to multiple-input and multiple-output systems is straightforward.\smallskip\\
A case study consisting of an electronic throttle body (ETB) \cite{b21} is considered to validate the proposed control strategies.\smallskip\\
The paper is organized as follows: Section \ref{sec:VRFT} shortly recalls the Virtual Reference Feedback Tuning method for the design of nonlinear SISO systems, Section \ref{sec:RNNs} shows the equations of the regulators defined by ESNs and LSTMs. Also, in Section \ref{sec:Control_Schemes} the advanced control scheme used for control of the ETB is presented. Finally, Section \ref{sec:ETB} discusses the application to the benchmark example, while conclusions are drawn in Section \ref{sec:conclusions}.

\section{Virtual Reference Feedback Tuning}
\label{sec:VRFT}
\begin{figure}[htbp]
	\centering
	\includegraphics[width=0.6\columnwidth]{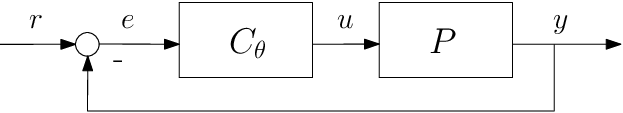}
	\caption{Control scheme}
	\label{fig: VRFTCS}
\end{figure}

In this section a brief overview on VRFT is given. VRFT (see, e.g., \cite{b5,b8,b9}) is a direct data-based controller design method, i.e., where a batch of data is commonly sufficient to obtain a controller and no iterations are necessary. Considering Figure \ref{fig: VRFTCS}, the objective of VRFT is to identify a controller $C_{\theta}$ (with unknown parameter vector $\theta$), such that the resulting closed-loop system is as similar as possible to a given reference closed-loop model \textit{M} for a given reference $r$ of interest.
This is done using, as available data, input and output sequences of length $N$ drawn from the plant. In this paper we consider the noise-free case, since in the considered case study noise is practically negligible (see Section \ref{sec:ETB}). If necessary, noisy data can be easily dealt with, e.g., using instrumental variable methods (as suggested in \cite{b9}).\\
The main idea behind VRFT is to compute the optimal value of the unknown controller parameter vector $\theta$ that solves the following problem:
\begin{alignat}{2}
\label{cost_gen}
\min \ J(\theta) := \|y_{\theta}-M[r]\|^{2} 
\end{alignat}
where $y_{\theta}(k+1) = P[C_{\theta}[r(k)-y_{\theta}(k)]]$ is the output sequence obtained by using the controller $C_{\theta}$ and $\|\cdot\|$ is the Euclidean norm. If $\{C_{\theta}, \theta\in{R}^{n_{\theta}}\}$ is sufficiently rich and $r$ is sufficiently exciting, solving \eqref{cost_gen} allows to obtain a controller that makes the closed-loop system as close as possible to \textit{M}. However, the optimization problem \eqref{cost_gen} cannot be formulated in practice, since the plant model $P$ is not known. Therefore, the so-called \textit{virtual reference} sequence $r$ consistent with the output sequence $y$ is computed according to
\begin{alignat}{2}
\label{VRFTinversion}
r = M^{-1}[y]
\end{alignat}
The sequence $r$ is denoted as \textit{virtual} because it does not exist in reality and it was not in place when data $u$ and $y$ were collected. It is indeed the reference signal that, if injected at the feedback control system input, it produces $y$ as output in accordance with $y = M[r]$. It is possible to define $e(k) := r(k)-y(k)$ as the \textit{virtual error} and to identify the optimal value of vector $\theta$ by minimizing the following alternative \textit{P}-free cost:
\begin{alignat}{2}
\label{cost_VRFT}
\min \ J_{VRFT}(\theta) := \|F[C_{\theta}[e]]-F[u]\|^{2}  
\end{alignat}
where $F: \mathbb{R}^{N}\to\mathbb{R}^{N}$ is a suitable filter. For simplicity, in the following we assume $F=I$. A good controller is one that produces $u$ when fed by $e$ because, through \textit{P}, this generates $y$, the desired output when reference is $r$.

The objective reference map \textit{M} can be selected as a general nonlinear system. In this work, in order to impose performances in a simple way, we use a linear system. 
We must select $M$ as invertible, in such a way that we can compute the sequence of reference values $r$ given the sequence of output values $y$ as in \eqref{VRFTinversion}.

Concerning the controller class $C_{\theta}$, it may be selected as a class of linear systems (e.g. PID controllers) or nonlinear ones. In this paper we will focus our attention on some notable classes of RNNs, in view of their notable properties, discussed in the following section.

\section{RNN regulators}
\label{sec:RNNs}
In this section the equations of the RNN regulators that will be used as controller class are defined, as well as the properties that can be guaranteed through suitable constraints on the unknown parameters.

\subsection{ESN regulators}
\subsubsection{Regulator equations}
Starting from the equations defined by Jaeger in \cite{b15}, it is possible to obtain the non-strictly proper version (to avoid to include an unnecessary delay in the computation of the control action) of the ESN equations, that will be used for the controller. We consider an ESN regulator with $n$ neurons (i.e. states $x$) in the reservoir, input $e$ and output $u$. Upstream we apply a suitable input scaling and shift thanks to parameters $k_e$ and $s_e$, respectively, i.e., 
\begin{subequations}\label{esnreg}
	\begin{align}
	\tilde{e}(k) = k_ee(k)+s_e \label{esnreg1}
	\end{align}	
The dynamics of the ESN is described by the equations
	\begin{align}
		& x(k+1) = g(\tilde{e}(k),x(k),\tilde{u}(k-1)) \label{esnreg2}\\
		& \tilde{u}(k) = W_{out_{1}}g(\tilde{e}(k),x(k),\tilde{u}(k-1))+W_{out_{2}}\tilde{e}(k) \label{esnreg3}
	\end{align}
where $g(\tilde{e}(k),x(k),\tilde{u}(k-1))= \tanh(W_{e}\tilde{e}(k)+W_{x}x(k)+W_{u}\tilde{u}(k-1))$ is the so-called \emph{activation function}. In the SISO case $W_{out_{1}}\in\mathbb{R}^{1\times n}$, $W_{out_{2}}\in\mathbb{R}^{1\times 1}$, $W_{e}\in\mathbb{R}^{n\times 1}$, $W_{x}\in\mathbb{R}^{n\times n}$, and $W_{u}\in\mathbb{R}^{n\times 1}$. Finally, the following downstream de-normalization is applied, thanks to scaling and shift parameters $k_u$ and $s_u$, respectively.
\begin{align}
	 u(k) = \frac{\tilde{u}(k)-s_u}{k_u} \label{esnreg4}
\end{align}
\end{subequations}
Importantly, the tuning of Echo State Networks is computationally advantageous over other types of RNNs. Indeed, while the choice of matrices $W_e$, $W_x$, and $W_u$ in the dynamic equation \eqref{esnreg2} is largely arbitrary (although they can be subject to optimizaton using, e.g., the approach developed in \cite{ScenarioESN}), matrices $W_{out_1}$ and $W_{out_2}$ are estimated based on a suitable least square-based optimization algorithm (see, e.g., \cite{b18}). Since the equation \eqref{esnreg3} is linear-in-the-parameters, the identification algorithm boils down to a computationally lightweight quadratic program. This, as a byproduct, makes it possible to straightforwardly apply instrumental variable methods when dealing with noisy data, as discussed in \cite{TesiWilliam}.
\subsubsection{Constrained inputs}
\label{sec:constrESN}
With reference to the ESN regulator \eqref{esnreg}, the following property holds.\smallskip\\
\textit{Property 1.} Consider an ESN regulator defined by equations \eqref{esnreg}, with $k_u>0$. If $W_{out_{2}} = 0$ in \eqref{esnreg3}, then
\begin{align}
\label{propertyesn}
\frac{-\|W_{out_1}\|_{\infty}-s_u}{k_u} \leq u \leq \frac{\|W_{out_1}\|_{\infty}-s_u}{k_u}
\end{align}
Recall that, given a matrix $M\in\mathbb{R}^{r\times c}$ with elements $m_{ij}$, $\|M\|_{\infty}$ is its $\infty$-norm, i.e., $\|M\|_{\infty}=\max_{i=1,\dots,r}\sum_{j=1}^c|m_{ij}|$. \smallskip\\
\begin{proof}
	Consider the output equation of an ESN regulator \eqref{esnreg3}. If $W_{out_{2}} = 0$, the equation becomes:
	\begin{align}
	\tilde{u}(k) = W_{out_{1}}\tanh(W_{e}\tilde{e}(k)+W_xx(k)+W_{u}\tilde{u}(k-1)) \notag
	\end{align}
	Since $\tanh(\cdot) \in (-1,1)$ element-wise,
	\begin{align}
	|\tilde{u}(k)| & = |W_{out_{1}}\tanh(W_{e}\tilde{e}(k)+W_xx(k)+W_{u}\tilde{u}(k-1))|\notag\\
	& \leq |W_{{out_{1}}_1}|+...+|W_{{out_{1}}_n}| = \|W_{out_{1}}\|_{\infty}	\notag
	\end{align}
	and so $-\|W_{out_{1}}\|_{\infty} \leq \tilde{u}(k) \leq \|W_{out_{1}}\|_{\infty}$. Recalling \eqref{esnreg4}, we obtain \eqref{propertyesn}.
\end{proof}	
\textit{Remark 1.} Property 1 can straightforwardly be extended to a MIMO regulator and in particular each component of the input vector is bounded in a similar way. The only difference lies in how inputs and outputs are normalized and de-normalized, since each element of the input and output vectors can be subject to different scaling operations.\medskip\\
Based on \eqref{propertyesn}, a regulator with a control variable $u$ constrained in a range [$l_b$,$u_b$] can be obtained as follows:
\begin{enumerate}
	\item Fix the output shift and scaling parameters, i.e.,
	\begin{align}\label{eq:scaling_esn}
	s_u = -\frac{u_b+l_b}{u_b-l_b} ,\,\,
	k_u = \frac{2}{u_b-l_b}
	\end{align}	
	\item Solve a convex constrained optimization problem, e.g. with YALMIP \cite{Yalmip}, i.e., 
	\begin{align}
	\underset{W_{out_1}}{\min}\ J := \|\tilde{u}-\hat{u}(W_{out_1})\|^2\label{eq:opt_esn}
	\end{align}
	subject to
	\begin{align*}
	\|W_{out_1}\|_{\infty} & \leq 1 \\
	W_{out_2} & = 0
	\end{align*}
	In \eqref{eq:opt_esn}, $\tilde{u}$ is the output data vector and $\hat{u}(W_{out_1})$ is the predicted output data vector, both normalized. 
\end{enumerate}
It is finally worth noting that, based on the results presented in \cite{b18}, it is possible to provide an analytical sufficient condition for guaranteeing that the ESN enjoys $\delta$-ISS \cite{b19b}, i.e., $\|W_x+W_uW_{out_1}\|<1$. The latter condition can be explicitly enforced in the optimization problem \eqref{eq:opt_esn} as a further constraint. In the experimental results illustrated in Section \ref{sec:ETB}, however, this condition turns out to be rather restrictive for the selection of suitable controllers. In view of this, we opted to remove this condition from the optimization problem, while checking \textit{a posteriori} suitable stability-related properties of the resulting regulator, as described later in the paper.
\subsection{LSTM regulators}
\subsubsection{Regulator equations}
Starting from the equations defined by Hochreiter $\&$ Schmidhuber in \cite{b16}, it is possible to obtain the nonstrictly proper version of the LSTMs. We define $\sigma_{g}(x) = \frac{1}{1+e^{-x}}$ and $\sigma_{c}(x) = \tanh(x)$ and, when applied to a vector, we assume to apply them element-wise. Furthermore, `$\circ$' represents the element-wise (Hadamard) product. 
We consider an LSTM regulator with $n_x$ hidden layers, input $e$ and output $u$. Upstream we apply the input scaling \eqref{esnreg1}.
The dynamics of the LSTM is described by the equations
\begin{subequations}\label{lstmreg}
	\begin{align}
	x(k+1) =& f(\tilde{e}(k),\xi(k)) \circ x(k)\notag\\ 
	&+i (\tilde{e}(k),\xi(k)) \circ a(\tilde{e}(k),\xi(k)) \label{lstmreg2}\\
	\xi(k+1) =& o(\tilde{e}(k),\xi(k)) \circ \sigma_c(f(\tilde{e}(k),\xi(k)) \circ x(k)\notag\\
	& + i(\tilde{e}(k),\xi(k)) \circ a(\tilde{e}(k),\xi(k))) \label{lstmreg3}\\
	\tilde{u}(k) = &W_{out}[o(\tilde{e}(k),\xi(k)) \circ \sigma_c(f(\tilde{e}(k),\xi(k)) \circ x(k)\notag\\ &+i(\tilde{e}(k),\xi(k)) \circ a(\tilde{e}(k),\xi(k)))]+b_{out} \label{lstmreg4}
	\end{align}
	where $f(\tilde{e},\xi) = \sigma_g(W_f \tilde{e} + U_f \xi + b_f)$ is the forget gate, $i(\tilde{e},\xi) = \sigma_g(W_i \tilde{e} + U_i \xi + b_i)$ the input gate, $a(\tilde{e},\xi) = \sigma_c(W_c \tilde{e} + U_c \xi + b_c)$ the input activation, $o(\tilde{e},\xi) = \sigma_g(W_o \tilde{e} + U_o \xi + b_o)$ the output gate, $W_{out} \in \mathbb{R}^{1,n_x}, b_{out} \in \mathbb{R}$, $W_f, W_i, W_o, W_c \in \mathbb{R}^{n_x,1}$, $U_f, U_i, U_o, U_c \in \mathbb{R}^{n_x,n_x}$, $b_f, b_i, b_o, b_c \in \mathbb{R}^{n_x,1}$. Finally, the downstream de-normalization \eqref{esnreg4} is used.\medskip\\
\end{subequations}
\subsubsection{Constrained inputs}
\label{sec:constrLSTM}
For LSTM regulators \eqref{lstmreg}, the following property holds.\smallskip\\
\textit{Property 2.} Consider an LSTM regulator defined by equations \eqref{esnreg1}, \eqref{lstmreg}, and \eqref{esnreg4}, with $k_u>0$. The output $u$ is bounded as follows
\begin{align}
	\label{propertylstm}
	\frac{-(\|W_{out}\|_{\infty}+|b_{out}|)-s_u}{k_u} \leq u \leq \frac{(\|W_{out}\|_{\infty}+|b_{out}|)-s_u}{k_u}
\end{align}

\begin{proof}
	Since, in view of their definitions, $\sigma_g(\cdot) \in (0,1)$ and $\sigma_c(\cdot) \in (-1,1)$ element-wise, then $\hat{\xi}\in (-1,1)$ as defined in equation \eqref{lstmreg3}. Hence, considering \eqref{esnreg4}, it is possible to write:
	\begin{align}
		|\tilde{u}(k)| & = |W_{out}\hat{\xi}+b_{out}| \notag\\
		& \leq \|W_{out}\|_{\infty}+|b_{out}|	\notag
	\end{align}
	and so $-(\|W_{out}\|_{\infty}+|b_{out}|) \leq \tilde{u}(k) \leq (\|W_{out}\|_{\infty}+|b_{out}|)$. Recalling \eqref{esnreg4}, we can write \eqref{propertylstm}.
\end{proof}	

It is worth noting that Remark 1 also applies to LSTM regulators, i.e., the extension to MIMO systems is straighforward.\medskip\\
Based on \eqref{propertylstm}, a regulator with a control variable $u$ constrained in a range [$l_b$,$u_b$] can be obtained as follows:
\begin{enumerate}
	\item Fix the output shift and scaling parameters according to \eqref{eq:scaling_esn}.
	\item Defining with $\theta$ the vector of tuning parameters, solve the following nonlinear constrained optimization problem 
	\begin{align}\label{eq:opt_lstm}
		\underset{\theta}{\min}\ J := \ \|\tilde{u}-\hat{u}(\theta)\|^2
	\end{align}
	subject to
	\begin{align*}
	\|W_{out}\|_{\infty} & \leq 1 \\
	b_{out} & = 0 
	\end{align*}
	where $\tilde{u}$ is the output data vector and $\hat{u}(\theta)$ is the predicted output data vector, both normalized.
\end{enumerate}		
Similarly to the case of ESNs, it is possible to provide an analytical sufficient condition for guaranteeing that the regulator enjoys $\delta$-ISS, i.e., that a suitably-defined matrix is Schur stable \cite{LSTM_stability}. Also in this case, being such condition only sufficient, it provided an overly-conservative constraint when used in \eqref{eq:opt_lstm}. For this reason in Section \ref{sec:ETB}, the stability-related properties of the LSTM are checked \textit{a posteriori}.
\section{Control scheme}
\label{sec:Control_Schemes}
\begin{figure}[h!]
	\centering
	\includegraphics[width=0.8\columnwidth]{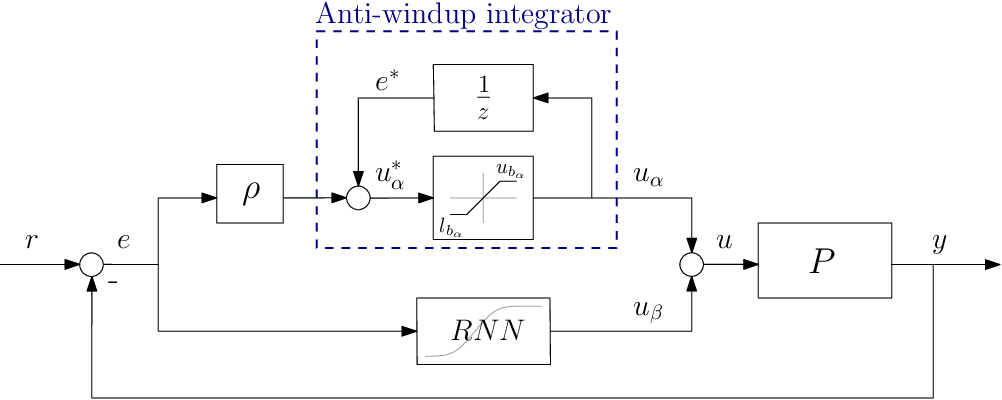}
	\caption{Anti wind-up integrator in parallel to the constrained RNN regulator}
	\label{fig: AntiParall}
\end{figure}

The advanced control scheme shown in Figure \ref{fig: AntiParall} will be used in the case study in order to achieve zero steady-state error and to bound the control variable avoiding the wind-up phenomenon. In particular, an anti wind-up integrator is placed in parallel (also when generating the data for the training of the RNN regulators) to the constrained ESN or LSTM regulator and, provided that the closed-loop system is asymptotically stable, a null steady-state error is guaranteed.

In this case, $u=u_\alpha + u_\beta$, where $u_\alpha$ and $u_\beta$ are the inputs generated by the integrator and the RNN, respectively. The gain $\rho$ of the integrator is chosen so as to take into account the different orders of magnitude of the error and the control variable. \\
The anti wind-up block is indeed used to effectively bound the input variable. If the control variable $u$ has to be bounded in a range [$l_b, u_b$], then the additive components $u_\alpha$ and $u_\beta$ must be bounded in such a way that $u$ is in the requested interval. In particular, we define $l_{b_\alpha}, l_{b_\beta}, u_{b_\alpha}, u_{b_\beta}$ in such a way that
\begin{align}
	l_b & = l_{b_\alpha} + l_{b_\beta} \notag \\
	u_b & = u_{b_\alpha} + u_{b_\beta} \notag
\end{align}	
and we will bound $u_\alpha$ (the integrator output) and $u_\beta$ (the RNN output) in a ``differential" way, i.e. by setting $u_\alpha\in[l_{b_\alpha}, u_{b_\alpha}]$ and $u_\beta\in[l_{b_\beta}, u_{b_\beta}]$. The definition of the bounds for $u_\alpha$ and $u_\beta$ is a design choice and depends on a trade-off: the wider the range of $u_\alpha$, the wider the range reachable by $u$ at steady-state but narrower the range of action $u_\beta$ of the RNN regulator which is exerted during the transient. Conversely, if a narrow range for $u_\alpha$ is chosen, the RNN regulator has a wider freedom of exerting its regulatory action but at steady-state $u$ turns out to be bounded in a narrow interval. \\
Since the limits of the saturation $l_{b_\alpha}$ and $u_{b_\alpha}$ are established before the training phase, we must keep in consideration the effect of the anti wind-up block also in the training phase. More in details, the training of the RNN is done by using $e$ as input and $u_\beta = u - u_\alpha$ as output, where $u_\alpha(k) = sat(u_\alpha^*(k)) =$
\begin{align*}
	\begin{cases}
		& l_{b_\alpha}\ ,\ u_\alpha^*(k) < l_{b_\alpha}\\
		& u_\alpha^*(k) = u_\alpha(k-1) + \rho e(k)\ ,\ l_{b_\alpha} \leq u_\alpha^*(k) \leq u_{b_\alpha}\\
		& u_{b_\alpha}\ ,\ u_\alpha^*(k) > u_{b_\alpha}
	\end{cases} \\
\end{align*}

\section{Experimental results}
\label{sec:ETB}

\subsection{System description and experimental setup}
\begin{figure}[h!]
	\centering
	\includegraphics[scale=0.35]{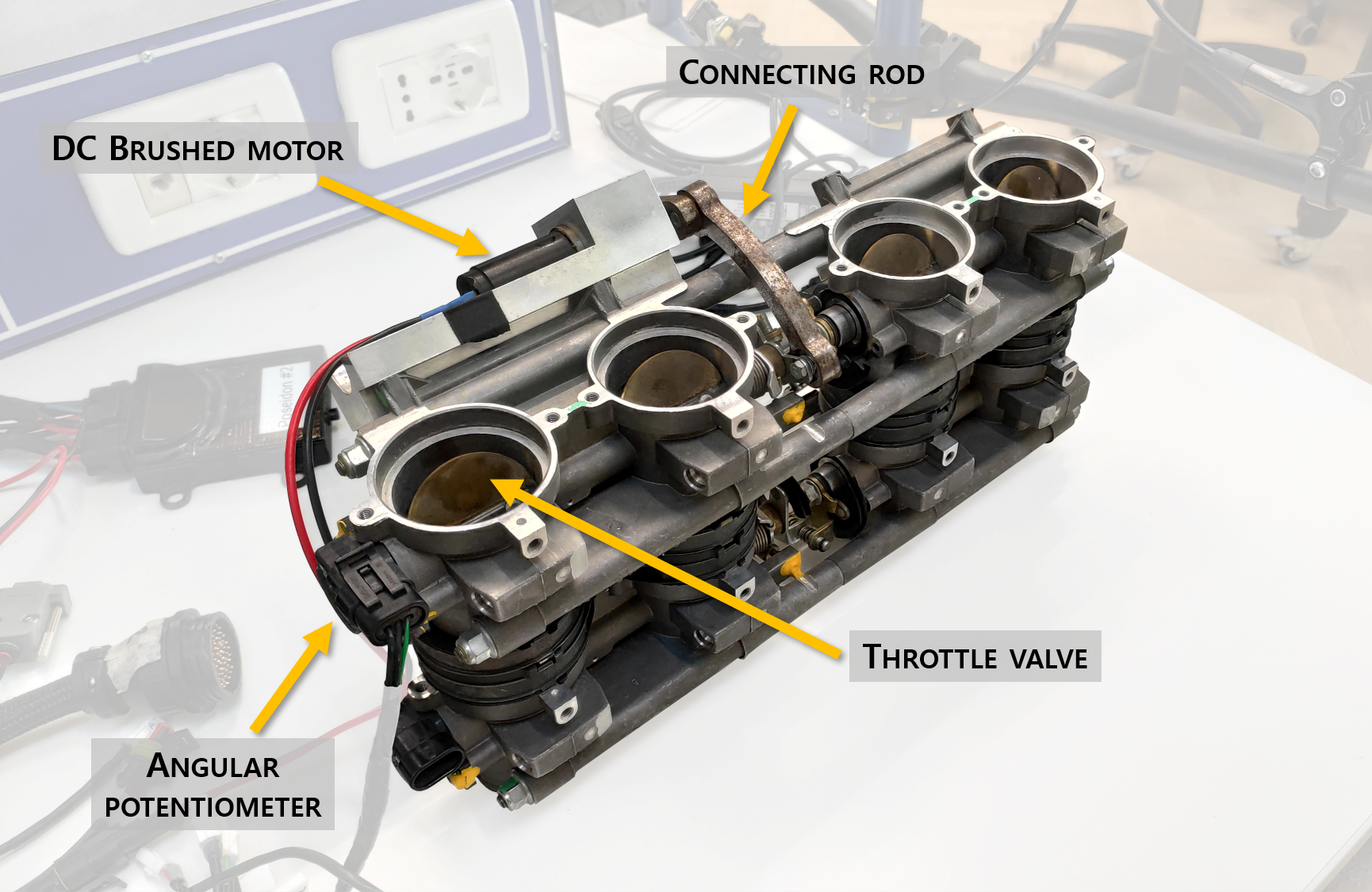}
	\caption{The electronic throttle body}
	\label{ThrottleSetup}
\end{figure}

The proposed controllers are validated on the electronic throttle body (ETB) depicted in Figure \ref{ThrottleSetup}.

The ETB is composed of four throttle valves actuated by a DC brushed motor, equipped with a planetary reduction gear and a connectiong rod which links the shaft of the motor to the shaft of the four valves. The input of the system is a voltage, obtained through a 20 kHz PWM signal, constrained in the range $[-12, 12]$ V. The output is the throttle shaft position measured by an angular potentiometer and defined in the range $[0, 1]$ (0 means that the valve is completely closed and 1 that it is totally open). 
The ETB is a nonlinear SISO system whose nonlinearities are mainly due to friction phenomena and to the nonlinear spring stiffness.

\subsection{Setting and available data}
\begin{figure}[h!]
	\centering
	\includegraphics[width=0.65\columnwidth]{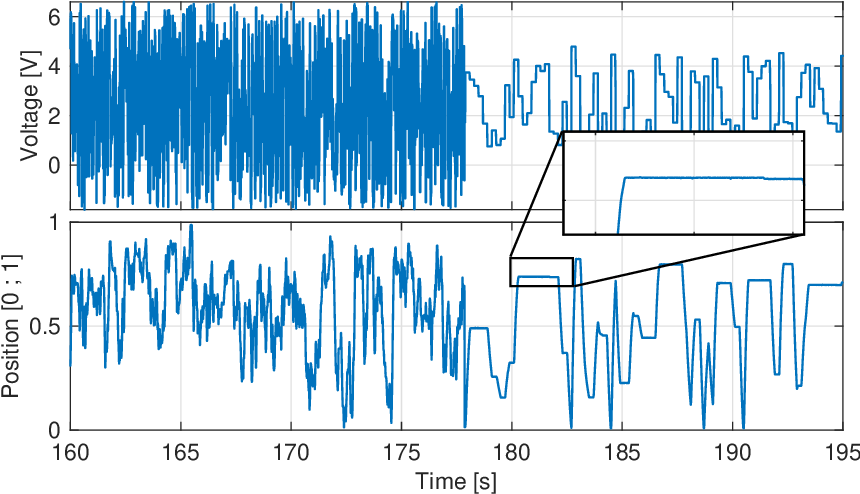}
	\caption{Portion of the data acquired from the ETB}
	\label{DataNoise}
\end{figure}

A sample time $T_s=0.005$ s is chosen for the data acquisition and the controller triggering. Since the settling time is equal to $10T_s$ when the maximum or minimum step of voltage is applied, a reasonable specification for the settling time of the closed-loop system is $20T_s$. This corresponds with the following unitary gain reference model: $y(k) = -a y(k-1) + b r(k-1)$, where we set $a=-0.79$ and $b=0.21$.\\
The available data are obtained with a Multilevel Pseudo-Random Signal (MPRS), \cite{b3}: half with switching period $4T_s$ and half with switching period $40T_s$. The amplitude of the PWM is in the range $[-1.8, 6.6]$ V so as to have the throttle shaft position covering the whole range of interest $[0, 1]$. Figure \ref{DataNoise} shows a portion of the data acquired from the ETB: note that, consistently with the assumption done in Section \ref{sec:VRFT}, noise on the data can be neglected.\\
\subsection{Controller identification}
\begin{figure}[h!]
	\centering
	\includegraphics[width=0.65\columnwidth]{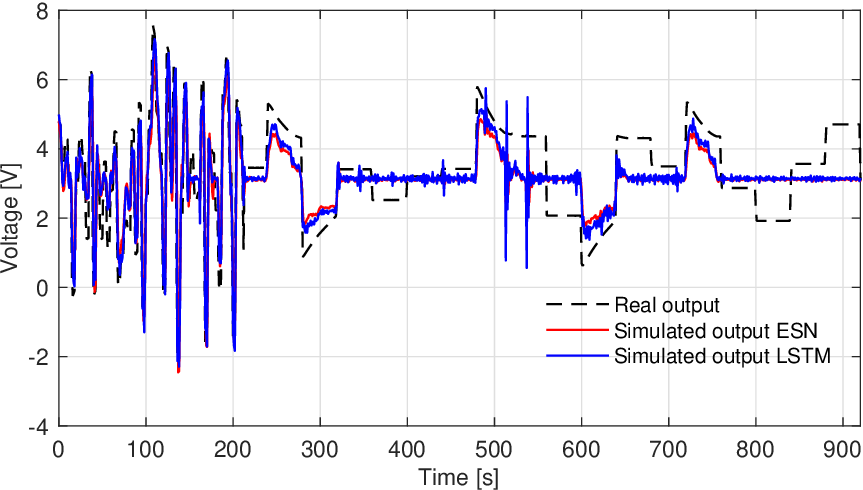}
	\caption{Portion of the validation data of ESN and LSTM}
	\label{DataFit}
\end{figure}

\begin{figure}[h!]
	\centering
	\includegraphics[width=0.65\columnwidth]{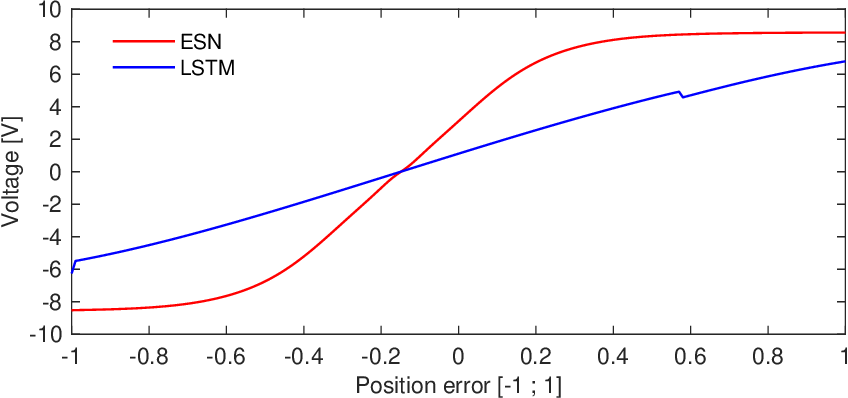}
	\caption{Static characteristic curves of the ESN and LSTM regulators}
	\label{Static}
\end{figure}

The applied control scheme is the one described in Section \ref{sec:Control_Schemes} with two alternative RNN regulators, i.e., a ESN and a LSTM. The gain $\rho$ of the integrator is set equal to $0.72$ and the saturation limits of the integrator are $-3.6$ V and $3.6$ V, allowing to reach all the throttle shaft positions in the range $[0,1]$ at steady-state. The ESN and LSTM regulators are bounded in the range $[-8.4,8.4]$ V: this makes it possible to bound the control variable $u$ in the range $[-12, 12]$ V.\smallskip\\
The training of the RNN regulators is carried out using $e$ as input and $u_\beta$ as output.\\
The ESN regulator is trained as in \cite{b18} setting $n=300$ according to the algorithm in Section \ref{sec:RNNs}, i.e. by means of the ESN learning toolbox developed by Jaeger \cite{b15} and by means of YALMIP \cite{Yalmip}.\\
The LSTM regulator is trained according to the algorithm in Section \ref{sec:RNNs} by means of the MATLAB function \textit{fmincon} and the implementation of the backpropagation through time (BPTT) algorithm, setting $n_x=25$. Inspired by the LSTM variants discussed in \cite{LSTM_variants}, and in order to lighten the computational burden in the training of LSTM in the constrained case, we have set $i(\tilde{e},\xi) = 1$ and $o(\tilde{e},\xi) = 1$ in such a way that the network in \eqref{lstmreg} takes the following simplified form:
	\begin{align*}
		x(k+1) =& f(\tilde{e}(k),\sigma_c(x(k))) \circ x(k) 
		+a(\tilde{e}(k),\sigma_c(x(k)))\\
		\tilde{u}(k) = &W_{out}[\sigma_c(f(\tilde{e}(k),\sigma_c(x(k)))\circ x(k)\notag\\  &+a(\tilde{e}(k),\sigma_c(x(k))))]
	\end{align*}
We also set $W_{out} = [1\ \ 0\ \ \cdots\ \ 0]$, naturally verifying $\|W_{out}\|_\infty \leq 1$ and $b_{out} = 0$. In this case the minimization is performed with respect to the matrices $W_f, W_c, U_f, U_c, b_f, b_c$ only.\\
The fitting parameter used to evaluate the quality of the identification procedure is
\begin{equation}
	\label{fit}
	FIT (\%) = 100\cdot\left(1-\frac{\|u-u_{sim}\|}{\|u-\bar{u}\|}\right)
\end{equation}
where $u$ is the acquired control variable, $u_{sim}$ is the control variable simulated by the network and $\bar{u}$ is the mean value of the acquired control variable $u$. In Figure \ref{DataFit}, a portion of the validation data of the ESN and LSTM regulators is depicted. 
The fitting value achieved by the ESN regulator is $51\%$, while the one obtained by the LSTM regulator is $55\%$.\\
Finally, as discussed in Sections \ref{sec:constrESN} and \ref{sec:constrESN}, the stability properties of the considered regulators are not imposed during the training phase. However, we have successfully numerically verified \textit{a posteriori} the asymptotic stability of the equilibria corresponding with the static characteristic curves of both regulators (see Figure \ref{Static}).
\subsection{Reference tracking results}
\label{sec:results01}
\begin{figure}[h!]
	\centering
	\includegraphics[width=0.65\columnwidth]{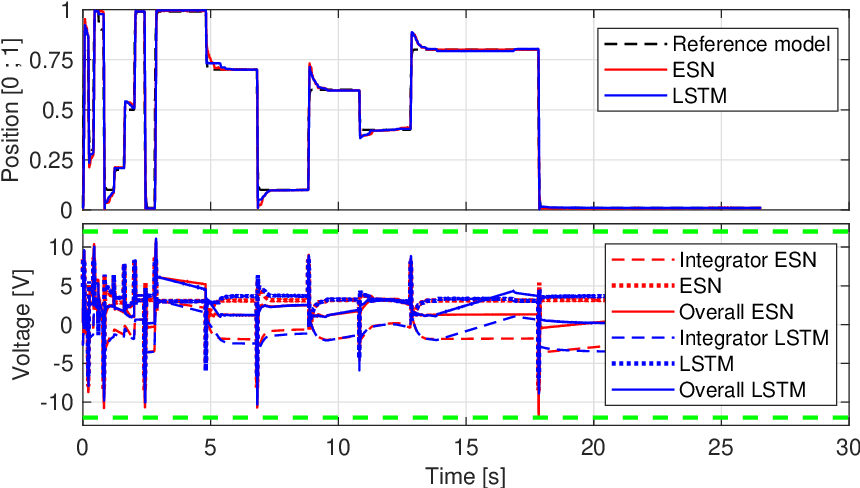}
	\caption{Reference tracking results}
	\label{Tracking1}
\end{figure}

\begin{figure}[h!]
	\centering
	\includegraphics[width=0.65\columnwidth]{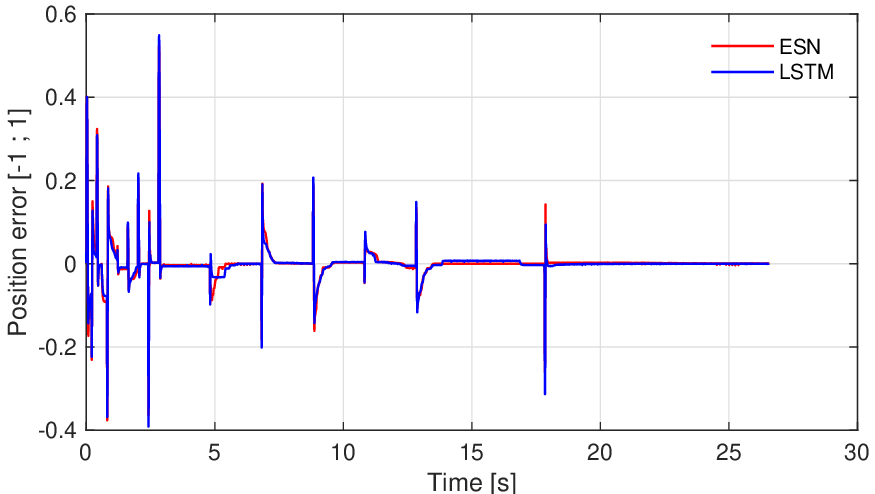}
	\caption{Reference tracking error}
	\label{Error1}
\end{figure}

A time-varying reference position signal is used in these tests. In Figure \ref{Tracking1}, the reference tracking results are shown for the considered RNN regulators, while for better insight the reference output error is shown in Figure \ref{Error1}. In the upper plot of Figure \ref{Tracking1}, the reference closed-loop model output is depicted with a black dashed line, the system output with an ESN regulator is displayed with a red line and the system output with a LSTM regulator with a blue line. The bottom plot shows $u_\alpha, u_\beta$, and $u$ when ESN and LSTM regulators are used, respectively. 
%
%
\subsection{Comparisons with different setups}
In this section the previous constrained RNN regulators are compared with other RNN regulators designed according to different design choices. These results are shown to motivate the adopted design choices, as well as the role of constraints in the tuning of the RNN and of the integrator.
\subsubsection{Different reference model}
\begin{figure}[h!]
	\centering
	\includegraphics[width=0.65\columnwidth]{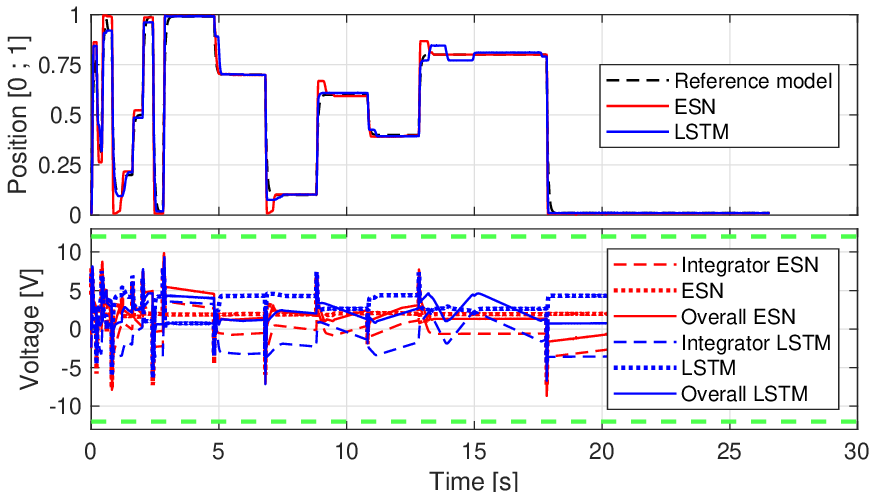}
	\caption{Reference tracking results with different reference model (settling time equal to $49T_s$)}
	\label{Tracking2}
\end{figure}

\begin{figure}[h!]
	\centering
	\includegraphics[width=0.65\columnwidth]{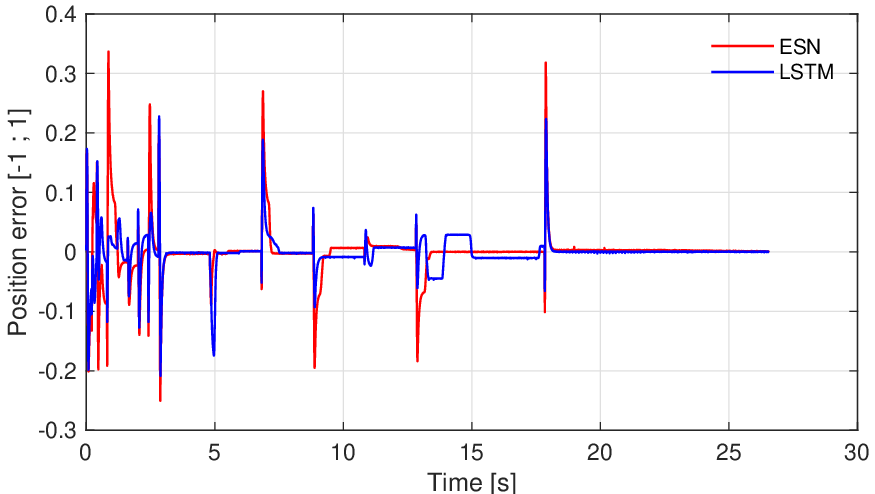}
	\caption{Reference tracking error with different reference model (settling time equal to $49T_s$)}
	\label{Error2}
\end{figure}

In Figure \ref{Tracking2} we show the reference tracking results of the RNN regulators when the reference closed-loop model has settling time equal to $49T_s$, i.e., where $a=-0.91$, $b=0.09$. In this case, the fitting index of the ESN regulator is equal to $47\%$ with $n=300$, while the one of the LSTM regulator is $30\%$ with $n_x=25$. It is possible to notice that the tracking performances are slightly worse with respect to the ones obtained in the case illustrated in Section \ref{sec:results01}. In particular, as apparent from Figure  \ref{Error2}, the closed-loop responses now display smaller overshoots with respect to the case in Section \ref{sec:results01}, but at the price of longer transients. \\

\subsubsection{Different integrator weight}
\begin{figure}[h!]
	\centering
	\includegraphics[width=0.65\columnwidth]{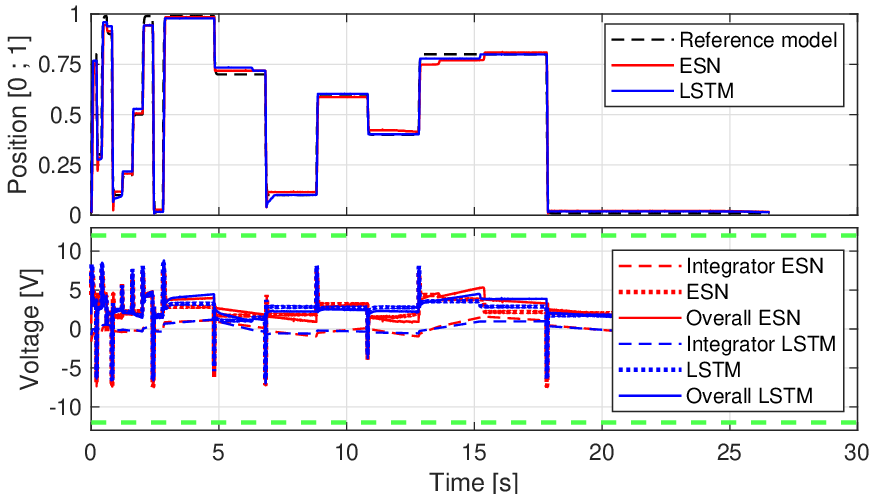}
	\caption{Reference tracking results with different integrator weight ($\rho=0.12$)}
	\label{Tracking3}
\end{figure}

\begin{figure}[h!]
	\centering
	\includegraphics[width=0.65\columnwidth]{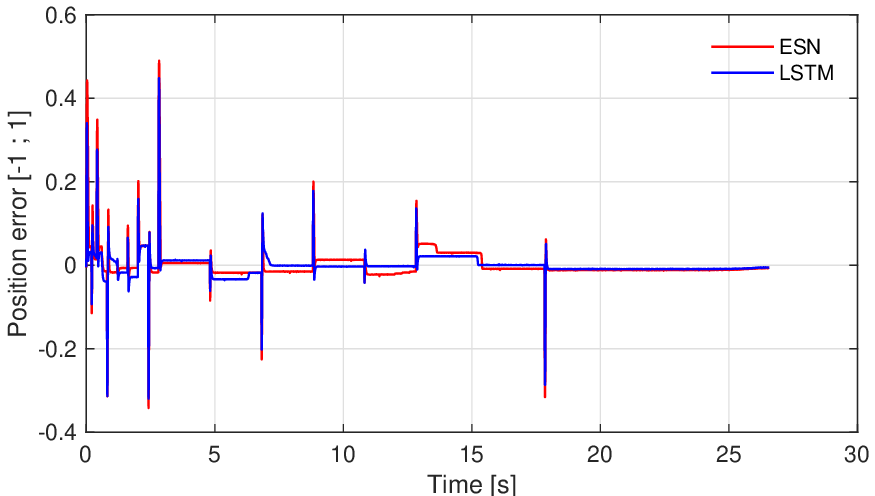}
	\caption{Reference tracking error with different integrator weight ($\rho=0.12$)}
	\label{Error3}
\end{figure}

Figure \ref{Tracking3} shows the reference tracking results of the constrained RNN regulators in parallel to a constrained integrator when the reference closed-loop model has settling time equal to $20T_s$ but the gain of the integrator is equal to $\rho=0.12$. The fitting of the ESN regulator is equal to $58\%$ with $n=300$, and the one of the LSTM regulator is $54\%$ with $n_x=25$. It is possible to notice from Figure \ref{Error3} that, due to the smaller gain of the actuator, the latter is less responsive than in the case considered in Section \ref{sec:results01}, where $\rho=0.72$. In view of this, the final transients of the closed-loop response are slower. \\

\subsubsection{Constrained network tuning and no integrator}
\begin{figure}[h!]
	\centering
	\includegraphics[width=0.65\columnwidth]{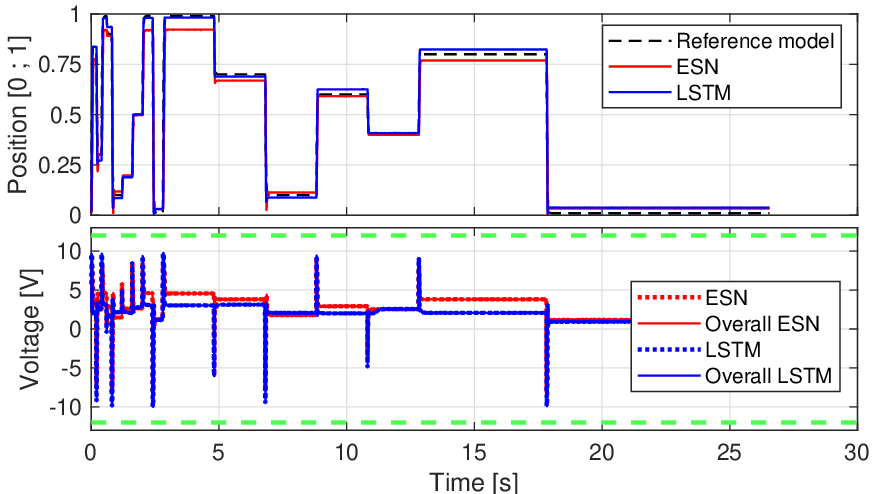}
	\caption{Reference tracking results with constrained network tuning and no integrator}
	\label{Tracking4}
\end{figure}

\begin{figure}[h!]
	\centering
	\includegraphics[width=0.65\columnwidth]{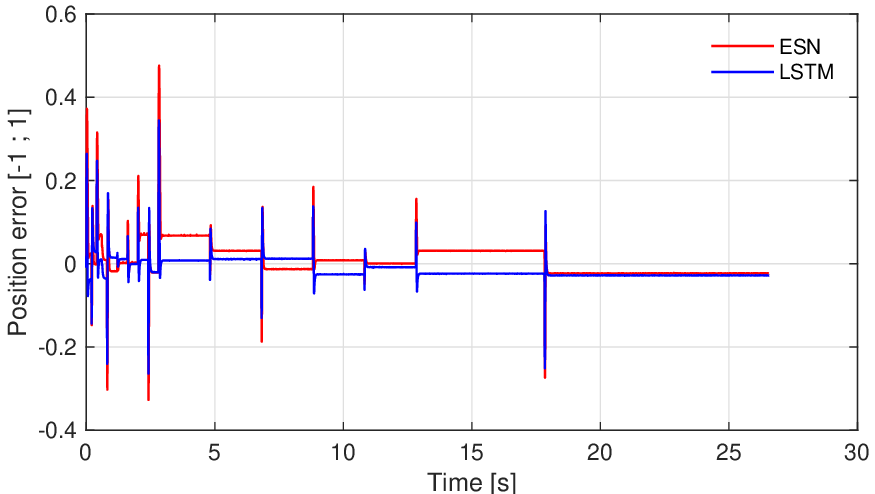}
	\caption{Reference tracking error with constrained network tuning and no integrator}
	\label{Error4}
\end{figure}

Figure \ref{Tracking4} shows the reference tracking results of the constrained RNN regulators applied without any integrator in the loop. The settling time of the reference closed-loop model is set to $20T_s$, the fitting of the ESN regulator is equal to $57\%$ with $n=300$, and the fitting of the LSTM regulator is equal to $51\%$ with $n_x=25$. Despite the tracking results in Figure \ref{Tracking4} appear to be rather promising, from an inspection of Figure \ref{Error4} we can observe that the static precision of the control system is not perfect (i.e., the responses display non-zero steady-state errors). This is due to the unavoidable regulator identification inaccuracies and highlights the importance of including suitable integral actions in the loop. \\
\subsubsection{Unconstrained network tuning and integrator}
\begin{figure}[h!]
	\centering
	\includegraphics[width=0.65\columnwidth]{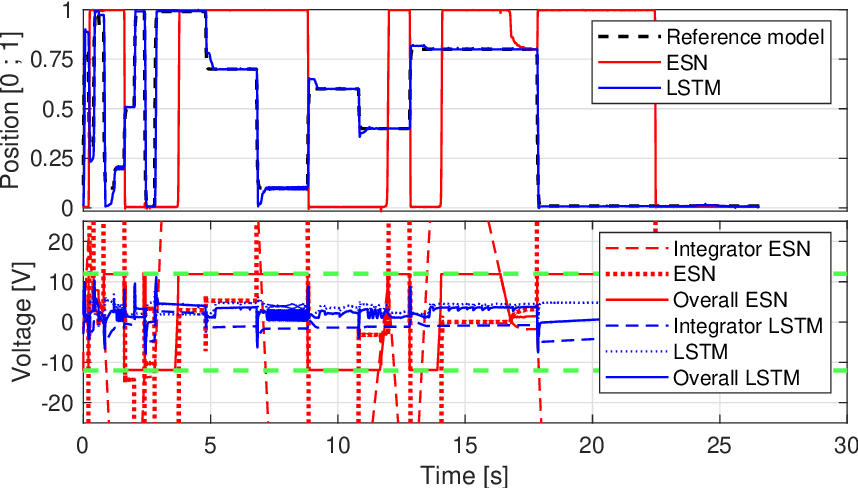}
	\caption{Reference tracking results with unconstrained network tuning and integrator ($\rho=0.72$)}
	\label{Tracking6}
\end{figure}
\begin{figure}[h!]
	\centering
	\includegraphics[width=0.65\columnwidth]{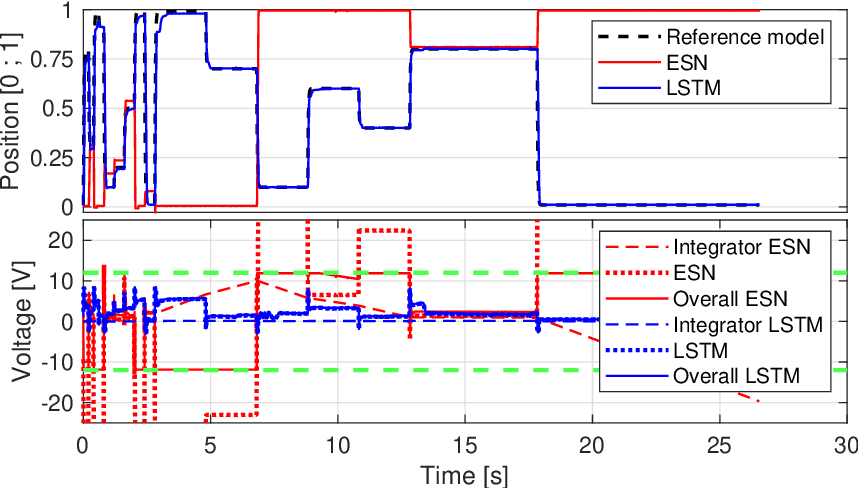}
	\caption{Reference tracking results with unconstrained network tuning and integrator ($\rho=0.012$)}
	\label{Tracking7}
\end{figure}
\begin{figure}[h!]
	\centering
	\includegraphics[width=0.65\columnwidth]{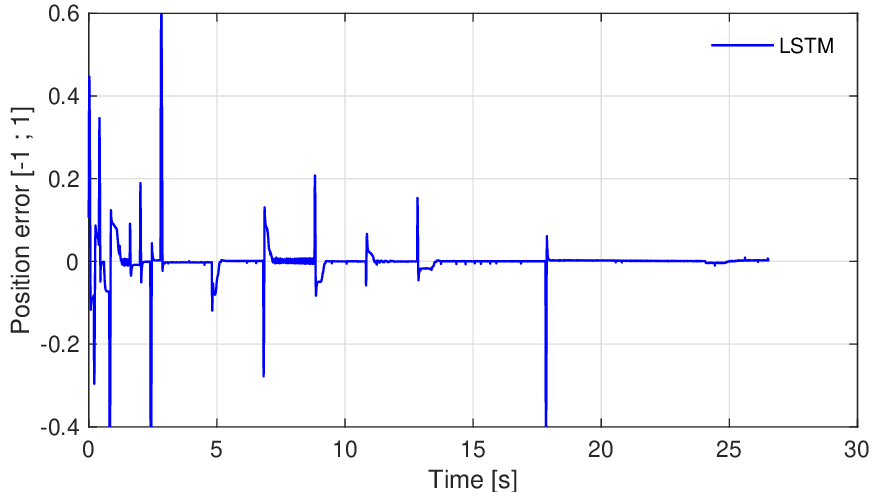}
	\caption{Reference tracking error with unconstrained network tuning and integrator ($\rho=0.72$)}
	\label{Error6}
\end{figure}
\begin{figure}[h!]
	\centering
	\includegraphics[width=0.65\columnwidth]{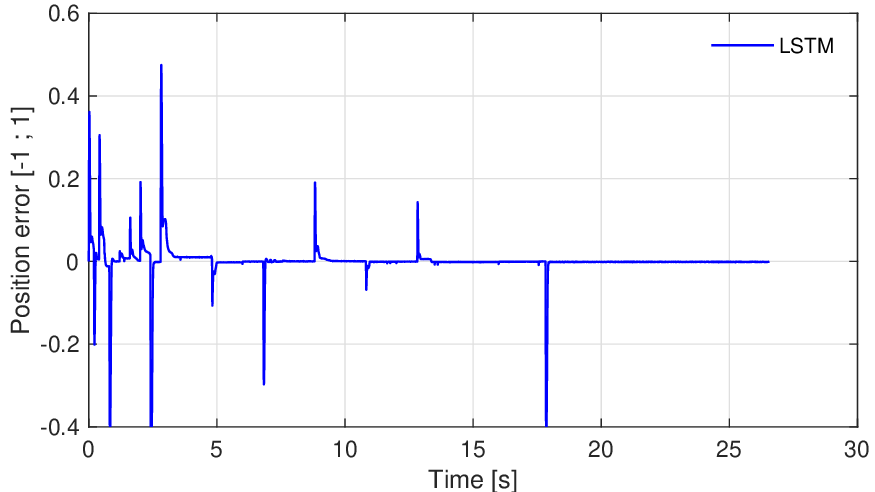}
	\caption{Reference tracking error with unconstrained network tuning and integrator ($\rho=0.012$)}
	\label{Error7}
\end{figure}

In Figures \ref{Tracking6} and \ref{Tracking7} the reference tracking results of the unconstrained RNN regulators applied with an unbounded integrator in the loop (with gain $\rho=0.72$ and $\rho=0.012$, respectively) are represented. The settling time of the reference closed-loop model is set to $20T_s$, the fitting of the ESN regulators with $n=300$ are equal to $62\%$ in both cases, and the fitting of the LSTM regulators with $n_x=25$ are equal to $85\%$ and $80\%$, respectively. In Figures \ref{Error6} and \ref{Error7} the reference output errors with the LSTM regulators are shown in the two considered cases, respectively. \\
In the first case, i.e. when $\rho=0.72$, it is possible to observe that several equilibria are not asymptotically stable, especially when the ESN regulator is applied. On the other hand, in the second case, i.e. when $\rho=0.012$, the response exhibits problems in reaching a zero steady-state error since the integral action is not sufficiently effective, especially with the ESN regulator. \\
However, the lack of control variable constraints makes the inclusion of an integral action in this scheme critical, as the integrator could generate indefinitely high values of the control variable and there are no guarantees that the output of the RNN regulators does not exceed the control variable limits. In particular, since the position error $e \in [-1, 1]$ in the considered case study, the output $u_\beta$ of the ESN regulator \eqref{esnreg} with $k_u>0$ is bounded as follows: 
\begin{align}
	\label{propertyesn2}
	\frac{-(\|W_{out_1}\|_{\infty}+|W_{out_2}|(|k_e|+|s_e|))-s_u}{k_u} \leq u_\beta \leq \frac{(\|W_{out_1}\|_{\infty}+|W_{out_2}|(|k_e|+|s_e|))-s_u}{k_u}
\end{align}
Hence, the output of the ESN regulator turns out to be bounded in $[-1.49\cdot 10^{8},  1.49\cdot 10^{8}]$ V if $\rho=0.72$ and in $[-1675.68, 1680.60]$ V if $\rho=0.012$. On the other hand, according to \eqref{propertylstm}, the LSTM regulator turns out to be bounded in $[-12.89, 19.05]$ V if $\rho=0.72$ and in $[-17.55, 22.39]$ V if $\rho=0.012$. Note that, indeed, the LSTM regulators perform better than the ESN ones. \\
\subsubsection{Unconstrained network tuning and no integrator}
\begin{figure}[h!]
	\centering
	\includegraphics[width=0.65\columnwidth]{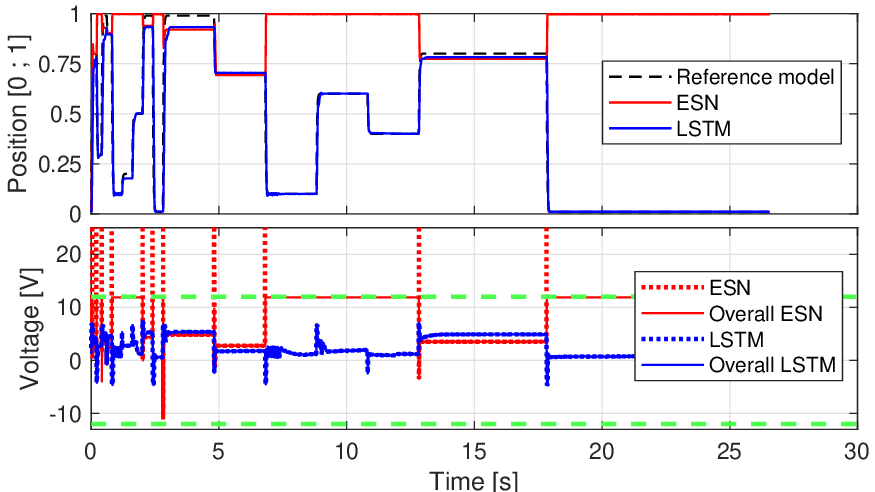}
	\caption{Reference tracking results with unconstrained network tuning and no integrator}
	\label{Tracking5}
\end{figure}

Finally, in Figure \ref{Tracking5} the reference tracking results of the unconstrained RNN regulators applied without any integrator in the loop are depicted. In this case the settling time of the reference closed-loop model is $20T_s$. The fitting of the ESN regulator is equal to $58\%$ with $n=300$, while the one of the LSTM regulator is $81\%$ with $n_x=25$. Regarding the ESN, the system response is highly unsatisfactory because the control variable often exceeds the saturation limits (the green dashed line represents the PWM limits). This underlines the practical importance of explicitly imposing input variable constraints in the regulator tuning. On the other hand, the response of the LSTM is less critical because it only exhibits problems in reaching a zero steady-state error when the values of the reference signal are close to 1. However, also in this case, there are no guarantees that the output of the RNN regulators does not exceed the control variable limits. According to \eqref{propertyesn2}, the ESN regulator is bounded in $[-1.73\cdot 10^{8}, 1.73\cdot 10^{8}]$ V. Instead, according to \eqref{propertylstm}, the LSTM regulator is bounded in $[-15.29, 20.11]$ V. \\
\section{Conclusions}
\label{sec:conclusions}
This paper describes the application of the VRFT method for control of nonlinear SISO systems with regulators described by ESNs and LSTMs. Their capability to constrain the control variable is pointed out and an advanced control scheme that allows to achieve a zero steady-state error is devised. Experimental results that validate the developed algorithms are shown considering a real ETB system. \\
Future works concern the comparison with other \textit{direct} or \textit{indirect} data-based methods and the extension of the developed control strategies to MIMO systems. Also, we will investigate how alternative choices of the reference model (including nonlinear ones) may affect the performances of the control systems.

\end{document}